# Recipes for Calibration Checks in Safety-Critical Applications

**Romeo Valentin**                                    ROMEOV@STANFORD.EDU

*Department of Aeronautics and Astronautics*
*Stanford University*
*Stanford, CA 94305, USA*

## Abstract

Safety-critical prediction systems, such as autonomous vehicles, weather forecasters, and medical monitors, commonly rely on probabilistic forecasters. These forecasters make predictions about possible future outcomes, and their quality and robustness needs to be validated and certified. Often, only accuracy — the mean of the predictions — is evaluated against true outcomes. However, for safety-critical scenarios and decision making under uncertainty, the full distributional properties of the forecasts should be checked: do the observed prediction errors actually follow the forecasted probability distributions? To this end, we introduce a *framework for calibration checks*: statistical tests that validate distributional properties of forecasts when measured over many samples. In order to support ease-of-use in real-world operations, these checks produce a single accept/reject decision for data collected from a forecaster. This contrasts typical calibration calculations which produce one or multiple continuous calibration scores and require expertise to implement in a validation workflow. We further support operationalization by introducing modifications to calibration testing that (a) reject only overconfident predictions, allowing for pessimistic or cautious predictions in safety-critical settings, and (b) tolerate small, operationally acceptable deviations even for large numbers of validation samples. We organize the calibration checking process into a modular pipeline comprising four steps: (i) the *data model*, (ii) the chosen *metric*, (iii) the *hypothesis formulation*, and (iv) the *testing procedure*. Each step consists of independently swappable components, thereby supporting a large variety of possible use-cases and trade-offs. We demonstrate the applicability of the framework on two complementary example problems, weather forecasting and robot pose estimation.

## 1 Introduction

A robot navigating near a cliff must know not only its position, but also how much to trust that position. A point estimate such as $\hat{x}_{\text{pos}} = 1.5\,\text{m}$ leaves no room for doubt: if the robot falsely believes it is safely away from the edge when it is in fact right on the edge, it drives over. The fix is an *uncertainty estimate*, a distributional state prediction such as $\hat{p}_{X_{\text{pos}}} = \mathcal{N}(1.5\,\text{m}, 0.3\,\text{m}^2)$ in place of a single number. Given the uncertainty, a planner can demand a wider safety margin whenever $\sigma$ is large, and only commit to the edge when it is small.

Uncertainty estimates are the foundation of robust decision making in safety-critical systems, and they come from two kinds of source. Classical systems including Kalman filters in aerospace, integrity monitoring in GNSS receivers (Blanch et al, 2015; Joerger and Pervan, 2016), and consistency analysis for SLAM (Bailey et al, 2006; Huang et al, 2010) derive their calibration guarantees from first principles. Learned components such as deep regressors, neural weather forecasters (Gneiting et al, 2005), and trajectory predictors for autonomous driving (Salzmann et al, 2020; Feng et al, 2021) have since vastly expanded the range of probabilistic outputs in deployed systems, but carry no analogous guarantee. For these systems, empirical *testing* is the only route to a calibration claim, and it is the subject of this paper.

Constructing such forecasters is its own subject; we assume one is given and focus on validating it. The core challenge is that a single forecast is typically paired with a single observation: if tomorrow's temperature falls far from the predicted mean, that does not on its own imply a bad forecaster, because even an ideal forecaster produces occasional long-tail errors. The fix is a frequentist reading over many samples: for a collection of 95% prediction intervals, 95% of outcomes should fall inside them (Dawid, 1984; Gneiting et al, 2007), at which point we may call the model *well calibrated at* 95% *coverage*.

This loose reading hides four design decisions that a deployment team has to settle before any check can be run. *What kind of prediction* does the system emit: a parametric density, a multivariate Gaussian with a covariance, a particle cloud from a filter? *How do we measure* the mismatch between forecasts and outcomes: by coverage of prediction intervals, or by the uniformity of transformed residuals? *What counts* as acceptable calibration: a two-sided match to the nominal level, a safety-critical one-sided bound that accepts benign conservatism, a tolerance budget that absorbs an engineering-specified miscalibration margin? And *when do we check*: offline on a fixed validation set, or online as predictions stream in during deployment? Each decision has been studied, but in largely separate literatures, and the literatures rarely speak to one another.

A further question cuts across all four. Most calibration evaluation in the machine-learning literature produces a *continuous score*, expected calibration error (Guo et al, 2017), negative log-likelihood, CRPS (Gneiting and Raftery, 2007), or a reliability diagram (Gneiting and Resin, 2023), that a human then has to threshold. A deployment specification cannot rest on "the number looks good": it needs a pass/fail verdict with a controlled false-rejection rate. We therefore frame calibration checking throughout as a *hypothesis test*, not a score.

This paper brings the four design decisions together into a single unifying framework, and makes each one's impact visible by treating it as an independently swappable slot: (i) the *data model*, which describes the form of forecasts and observations; (ii) the *metric*, which quantifies the discrepancy between forecasts and outcomes; (iii) the *hypothesis formulation*, which turns calibration into a testable statement; and (iv) the *testing procedure*, which produces the final accept/reject decision. The slot structure lets the framework absorb results from communities that have developed largely in parallel: per-level coverage from machine learning and conformal prediction, the probability integral transform from forecast verification and Bayesian computation, rank histograms from ensemble weather prediction, consistency analysis from robotics and SLAM, equivalence testing from biostatistics, and anytime-valid e-values from sequential analysis. Throughout, we adopt a safety-critical perspective: every recipe drawn from the framework yields a single accept/reject decision, can be made one-sided so that overconfidence is rejected while benign conservatism is accepted, and admits an explicit tolerance for operationally irrelevant deviations.

**Contributions.**

1. **A modular framework for regression calibration testing.** A four-slot decomposition (*Model & Data*, *Metric*, *Hypothesis*, *Testing*) organises the probabilistic-forecast evaluation literature into a single grammar whose slots are independently swappable. Every recipe drawn from the grammar returns a single accept/reject decision rather than a continuous score that still needs human judgement to operationalise. This decomposition is the primary contribution and the vehicle for (2) and (3).

2. **Safety-critical modifications.** We introduce two refinements that promote a generic calibration test to a deployable specification: (i) one-sided formulations that reject only overconfidence and accept benign conservatism, and (ii) tolerance bands that admit small, operationally irrelevant deviations within an explicit engineering budget. The constituent techniques are drawn from prior literature; our contribution is their consistent integration across every slot of the grammar, so that either refinement composes freely with any metric, model, or testing choice.[1]

3. **Empirical validation of composability.** We instantiate the framework on two safety-critical problems, a univariate Gaussian weather forecaster and a multivariate particle-filter localizer, each receiving a recipe that differs from the other on every slot: the weather recipe is a two-sided offline KS test on PIT uniformity (parametric model, offline *p*-value), and the robot recipe is a one-sided online e-value monitor on half-plane projections of the particle cloud (non-parametric model, online likelihood-ratio martingale). An appendix of four slot-swap ablations (Appendix A) holds the data fixed and varies one slot at a time, demonstrating that recipes drawn from the same grammar yield operationally distinct, engineer-readable decisions on identical data.

**Roadmap.** Section 2 fixes notation. Section 3 walks through the framework's simplest instance end-to-end (univariate Gaussian, per-level coverage, two-sided hypothesis, Bonferroni). The four refinement chapters then each vary one slot. They do not follow the pipeline's left-to-right order: we refine the *metric* first (Section 4), then the *hypothesis* (Section 5), then the *model & data* (Section 6), and finally the *testing* slot (Section 7). The ordering is pedagogical: the folded PIT developed in the metric slot is the object that lets the one-sided KS test of the hypothesis slot be stated cleanly, and once the two inner slots are in place the outer slots admit near-mechanical generalisation. Section 8 then validates the framework on two recipes at opposite corners of the design space, and Appendix A (appendix) isolates each slot in turn on the same data. Section 9 positions the framework against prior work.

## 2 Preliminaries

Uppercase letters denote random variables and lowercase their realisations: $Y_i$ is the $i$-th outcome, $y_i$ its value; the convention carries over to the PIT variables $U_i, V_i$ and $u_i, v_i$ of

---

[1] Prior-art ingredients: one-sided hypotheses (classical), the folding transformation from MCMC diagnostics (Vehtari et al, 2021) and simulation-based calibration (Modrák et al, 2023), tolerance bands from equivalence testing (Schuirmann, 1987; Lakens, 2017), the half-plane / sliced-Wasserstein construction (Rabin et al, 2012; Bonneel et al, 2015), and sequential e-values from safe testing (Henzi and Ziegel, 2022; Casgrain et al, 2024).

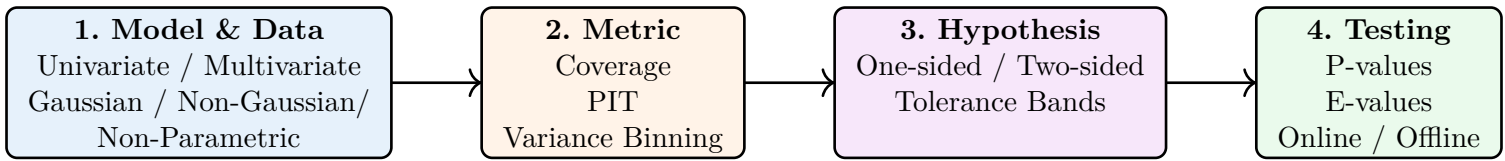

Fig. 1: The four-step calibration validation pipeline.

Section 4. A prediction model $f$ maps inputs $x_i$ to a predicted distribution $\hat{p}_{Y_i}(x_i) := f(x_i)$; the hat marks it as an estimate. The unknown true conditional distribution is $p^*_{Y_i}(x_i)$. The validation set is $N$ tuples $\{(x_i, \hat{p}_{Y_i}, y_i)\}_{i=1}^{N}$. Distribution objects are queried in function-call form: $\mathrm{cdf}(\hat{p}_{Y_i}, y_i)$, $\mathrm{qtl}(\hat{p}_{Y_i}, \alpha)$, and $\mathrm{pdf}(\hat{p}_{Y_i}, y_i)$.

Following the calibration hierarchy of (Dawid, 1984; Gneiting et al, 2007), the strongest notion is *conditional calibration* ($\hat{p}_{Y_i} = p^*_{Y_i}$ pointwise), which is untestable from a finite sample. We target the weaker, testable *probabilistic calibration*: the PIT variables $U_i = \mathrm{cdf}(\hat{p}_{Y_i}, Y_i)$ are i.i.d. $\mathcal{U}(0,1)$. Strict i.i.d. holds when the $(x_i, Y_i)$ pairs are independent; in the sequential settings of Section 8 it is only approximate, but temporal dependence is typically weak enough that the uniformity tests remain valid. We alternate freely between the language of *coverage* (the fraction of $y_i$ inside a level-$\lambda$ prediction set should equal $\lambda$) and *PIT uniformity*; Section 4.2 shows that the two are windows on the same condition.

## 3 The Main Pipeline

Calibration testing splits cleanly into four sequential design questions:

1. **Model & Data:** What type of predictions does the system produce?

2. **Metric:** How do we measure agreement between predictions and outcomes?

3. **Hypothesis:** What counts as acceptable calibration?

4. **Testing:** How do we turn that measurement into a decision?

Each question is a slot in the pipeline of Fig. 1; a *recipe* is a particular choice of one option per slot. Different applications call for different recipes, and the slots swap independently: changing the data model (e.g., Gaussian to particles) leaves the hypothesis and testing slots untouched, and trading offline p-values for online e-values does not require re-deriving the metric. Our aim throughout is to decide, from a finite validation sample, whether the predicted distributions are dangerously over-confident.

The rest of this section walks through the simplest concrete recipe (univariate Gaussian model, per-level coverage, a two-sided hypothesis test, and Bonferroni-corrected p-values), making the framework concrete before each slot is refined in turn.

### 3.1 Step 1: Univariate Gaussian Predictions

Fix the Model & Data slot to the univariate Gaussian: at each step $i$,

$$\hat{p}_{Y_i}(x_i) := \mathcal{N}(\mu_i, \sigma_i^2), \tag{1}$$

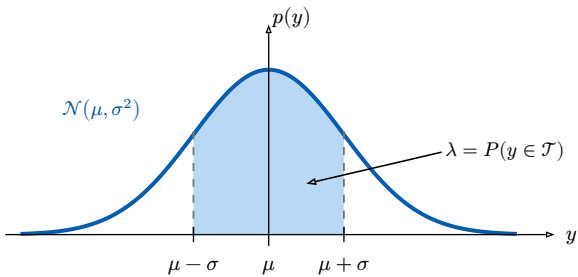

Fig. 3: A coverage set $\mathcal{T}_{0.68}(\mathcal{N}(\mu,\sigma))$ per (2).

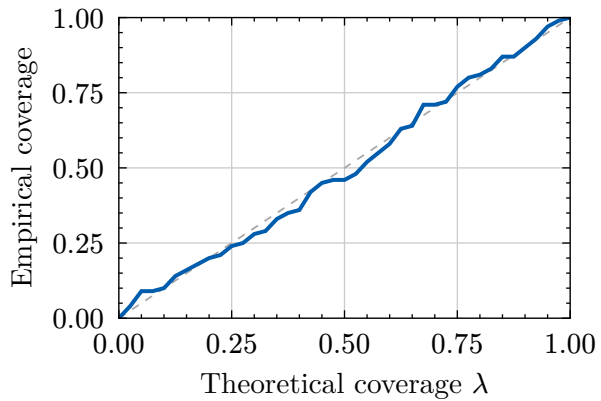

Fig. 4: Empirical coverage versus the ideal diagonal for a sweep of $\lambda$.

where $\hat{p}_{Y_i}$ is the model's predicted conditional distribution for the outcome $Y_i$, and the true distribution $p^*_{Y_i}$ may or may not be Gaussian. The univariate Gaussian is the ideal starting point: its CDF, quantile function, and prediction sets all admit closed forms, so every downstream step (counting, Binomial testing, Bonferroni correction) stays analytic. Other prediction types (multivariate, non-Gaussian, non-parametric) replace only this slot; the metric, hypothesis, and testing slots remain unchanged (Section 6).

**3.2 Step 2: The Coverage Metric**

For a level $\lambda \in (0,1)$, the *centered prediction set* at level $\lambda$ is the interval carrying probability mass $\lambda$ symmetrically around the median:

$$\mathcal{T}_\lambda(\hat{p}_{Y_i}) = \left[\mathrm{qtl}(\hat{p}_{Y_i},(1-\lambda)/2),\quad \mathrm{qtl}(\hat{p}_{Y_i},(1+\lambda)/2)\right]. \tag{2}$$

For a Gaussian this recovers the familiar symmetric intervals: $\mathcal{T}_{68\%}(\mathcal{N}(\mu,\sigma^2)) \approx [\mu-\sigma, \mu+\sigma]$ and $\mathcal{T}_{95\%}(\mathcal{N}(\mu,\sigma^2)) \approx [\mu-2\sigma, \mu+2\sigma]$ (Fig. 3). We use centered rather than highest-posterior-density (HPD) sets: the two coincide for symmetric distributions, and for skewed distributions centered sets are *more conservative* (they include the thin tail), which is the safe direction. Centred sets also carry one more benefit: they connect cleanly to the folded PIT in Section 4.2.

If the model is *probabilistically calibrated*, the fraction of outcomes falling inside $\mathcal{T}_\lambda(\cdot)$ equals $\lambda$ exactly (Christoffersen, 1998; Gneiting et al, 2007):

$$\mathbb{P}_{Y \sim p^*_Y}(Y \in \mathcal{T}_\lambda(\hat{p}_Y)) = \lambda \quad \forall \lambda \in (0,1). \tag{3}$$

This is the working definition of calibration in the ML literature (Kuleshov et al, 2018), and the most operationally interpretable one: "the 95% prediction interval should contain 95% of outcomes" maps directly onto an engineering specification. In practice we check (3) on a finite grid $\Lambda = \{\lambda_1, ..., \lambda_K\} = \{0.05, 0.15, ..., 0.95\}$ and plot the empirical against the nominal level (Fig. 4), the regression analogue of the classification *reliability diagram* (Gneiting and Resin, 2023). The discretisation introduces a multiple-testing problem, which Step 4 handles.

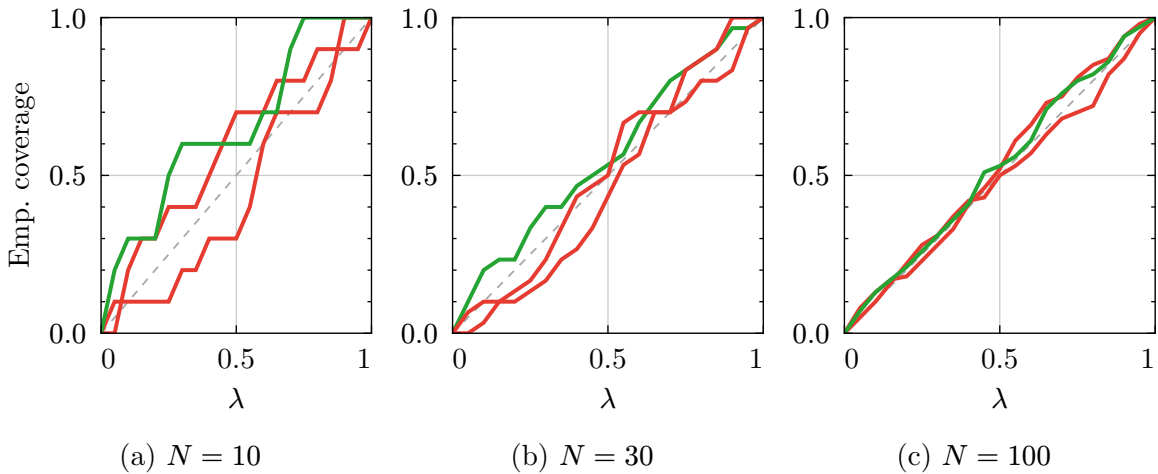

(a) $N = 10$  (b) $N = 30$  (c) $N = 100$

Figure 5: Empirical coverage curves from a perfectly calibrated model ($\hat{p} = p^* = \mathcal{N}(0,1)$) at three sample sizes, with three independent draws each. Green: curve lies entirely at or above the diagonal (appears conservative). Red: curve dips below at some level (appears over-confident). Even a perfect model can appear over-confident at small $N$.

### 3.2.1 Finite-sample fluctuation

Even with $\hat{p} = p^*$, the empirical coverage curve wanders off the diagonal in finite samples. Figure 5 shows three independent draws from a perfectly calibrated model at three sample sizes. At $N = 10$ the curves fluctuate wildly: some sit entirely above the diagonal (the model appears conservative by chance), others dip below (it appears over-confident by chance). The curves tighten around the diagonal as $N$ grows but never fall exactly onto it. Quantifying this random variation is the job of the Binomial model in Step 3.

### 3.3 Step 3: Hypothesis test

Let

$$c_{\lambda_k} = \sum_{i=1}^{N} \mathbb{1}\left[y_i \in \mathcal{T}_{\lambda_k}(\hat{p}_{Y_i})\right] \tag{4}$$

be the observed coverage count at level $\lambda_k$. Under calibration,

$$H_0 : c_{\lambda_k} \sim \text{Binomial}(N, \lambda_k), \tag{5}$$

and the natural finite-sample-valid test is two-sided: the count is rejected when it is either too low (intervals are too narrow on average, an over-confident model) or too high (intervals are too wide, a conservative model),

$$\begin{aligned}\text{Reject } H_0 \text{ at level } \tau_p \quad \text{iff} \quad & c_{\lambda_k} < \text{qtl}(\text{Binomial}(N, \lambda_k), \tau_p/2) \\ \text{or} \quad & c_{\lambda_k} > \text{qtl}(\text{Binomial}(N, \lambda_k), 1 - \tau_p/2),\end{aligned} \tag{6}$$

where $\text{qtl}(\cdot, \alpha)$ is the lower quantile (the smallest $c$ with $\mathbb{P}(C \leq c) \geq \alpha$). This is the classical two-sided Binomial test and makes no distinction between over- and under-coverage. Section 5 returns to this hypothesis slot and refines it along two safety-critical axes: a

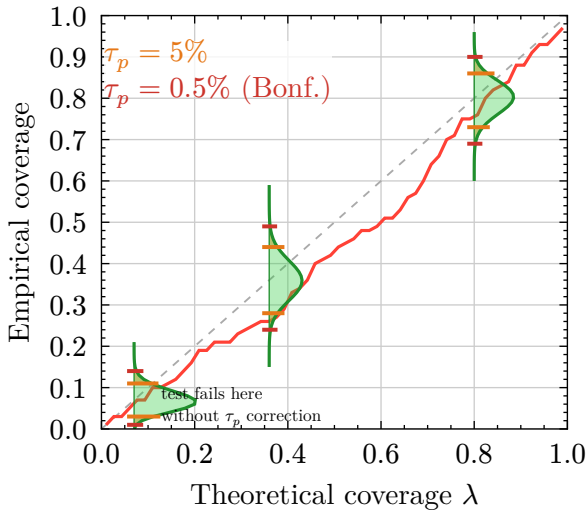
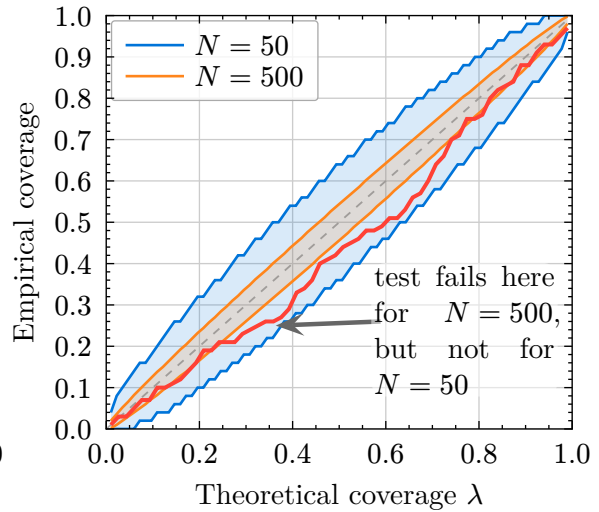

Fig. 7: Distribution of empirical coverage counts (green densities) at $\lambda \in \{0.07, 0.36, 0.80\}$ with $N = 100$ samples. The two-sided rejection region (shaded) lies in both tails of the Binomial at two thresholds: $\tau_p = 5\%$ (orange) and the Bonferroni-corrected $\tau_p = 0.5\%$ (red). A single empirical coverage trace (red) illustrates the stochastic variability.

Fig. 8: Two-sided coverage bounds at $\tau_p = 0.05$ for $N \in \{50, 500\}$. As $N$ grows, the bounds tighten. An empirical coverage curve (red) stays within the $N = 50$ bounds but exits the $N = 500$ bounds, illustrating the large-$N$ problem (Section 5).

*one-sided* variant that accepts conservatism as non-dangerous, and a *tolerance* variant that admits small miscalibration within an engineering-specified margin.

**3.4 Step 4: P-values and Multiple Testing**

The p-value at level $\lambda_k$ is the probability of a count at least as low as the observed one under $H_0$:

$$p_k = \mathbb{P}\big(c \leq c_{\lambda_k} \mid c \sim \text{Binomial}(N, \lambda_k)\big), \tag{7}$$

and the individual test rejects when $p_k < \tau_p$.

Running $K = |\Lambda|$ such tests simultaneously inflates the family-wise error rate: under independence the probability that *at least one* test falsely rejects is $1 - (1 - \tau_p)^K$, or roughly 40% for $K = 10$ and $\tau_p = 0.05$, and the fact that the coverage tests are positively correlated (they share samples) pulls the true rate down a little but leaves it far above $\tau_p$. The simplest remedy is the Bonferroni correction (Dunn, 1961): replace $\tau_p$ with $\tau_p/K$ in each individual test, so that the decision rule becomes

$$\text{accept at level } \tau_p \quad \text{iff} \quad \bigwedge_{\lambda_k \in \Lambda} p_k \geq \tau_p/K, \tag{8}$$

and the family-wise error rate is controlled regardless of dependence. The Holm–Bonferroni step-down procedure (Holm, 1979) sharpens this slightly, but neither exploits the specific

---

1 **Input:** predictions $\{(\mu_i, \sigma_i)\}$, outcomes $\{y_i\}$, levels $\Lambda = \{\lambda_1, ..., \lambda_K\}$, significance $\tau_p$
2 $\tau'_p \leftarrow \tau_p/K$ ▷ Bonferroni correction
3 **for** $\lambda_k \in \Lambda$ **do**
4 $\quad$ lo$_i \leftarrow$ qtl$(\mathcal{N}(\mu_i, \sigma_i^2), (1-\lambda_k)/2)$ for each $i$
5 $\quad$ hi$_i \leftarrow$ qtl$(\mathcal{N}(\mu_i, \sigma_i^2), (1+\lambda_k)/2)$ for each $i$
6 $\quad c \leftarrow \sum_{i=1}^{N} \mathbb{1}[y_i \in [\text{lo}_i, \text{hi}_i]]$
7 $\quad$ **if** $c <$ qtl$(\text{Binomial}(N, \lambda_k), \tau'_p/2)$ **or** $c >$ qtl$(\text{Binomial}(N, \lambda_k), 1-\tau'_p/2)$ **then**
8 $\quad\quad$ | **return** reject ▷ miscalibrated at this level
9 $\quad$ **end**
10 **end**
11 **return** pass

---

Recipe 1: **Univariate Gaussian two-sided coverage check with Bonferroni.** The simplest instantiation of the four-slot pipeline of Section 3 (univariate Gaussian model, coverage metric, two-sided hypothesis, Bonferroni-corrected $p$-values). Used as the pedagogical reference point for all subsequent slot swaps; the ablation of Appendix A.1 exercises this recipe on the weather data.

correlation structure of coverage tests; Section 4.2 returns to this point, showing via a thought experiment in incremental conditioning that the right joint test in the dense-grid limit is the Kolmogorov–Smirnov test on PIT values (Gneiting et al, 2007).

### 3.5 Putting the pieces together

The result is a calibration check that is *interpretable* (each level has a direct probabilistic meaning), *rigorous* (Bonferroni controls the false-alarm rate), and *simple* (counting, Binomial quantiles, comparison). We refer to this concrete instantiation of the four-slot pipeline as Recipe 1, and treat it as the pedagogical reference recipe throughout the rest of the paper: every refinement in Section 4, Section 5, Section 6, and Section 7 is described as a swap of one or more slots relative to this baseline, and the experiments of Section 8 instantiate two further recipes that jointly exercise all four swaps.

The basic pipeline has three visible limitations and one missing capability that the rest of the paper addresses slot by slot. Bonferroni ignores the positive correlation across levels and pays for it in statistical power (Section 4); the two-sided hypothesis rejects conservatism and over-confidence equally, and the exact null admits no slack for benign model mismatch, so the pipeline is not yet safety-critical and suffers from the *large-N problem* (Section 5); the univariate Gaussian assumption restricts the model family (Section 6); and the offline batch paradigm fixes $N$ up front, precluding runtime monitoring (Section 7).

### 4 Refining the Metric

The Metric slot turns out to offer two equivalent views of probabilistic calibration (Gneiting et al, 2007): per-level coverage and PIT uniformity. Section 4.1 introduces the probability integral transform (PIT), shows that the empirical coverage curve is a reparametrisation of the empirical CDF of folded PIT values, and presents a single Kolmogorov–Smirnov test on the folded PIT as a joint replacement for the $K$ Bonferroni-corrected per-level Binomials. Section 4.2 then explains why that replacement recovers the statistical power the coverage view leaves on the table, via a thought experiment on correlation-aware family-wise correction whose dense-grid limit is the Brownian bridge. Section 4.3 wraps any of these tests with binning by predicted uncertainty, turning a marginal check into a conditional one.

**4.1 The PIT as an Alternative Metric**

The coverage tests of Section 3 check the curve at $K$ discrete points and pay a Bonferroni penalty for doing so. There is a simpler route: test the joint calibration condition directly with a single uniformity check on transformed values.

The probability integral transform (PIT) for sample $i$ is

$$U_i = \text{cdf}(\hat{p}_{Y_i}, Y_i). \tag{9}$$

Under probabilistic calibration $U_i \sim \mathcal{U}(0,1)$ i.i.d. (Rosenblatt, 1952; Dawid, 1984; Diebold et al, 1998), and the full calibration condition collapses to the single statement that the $u_i = \text{cdf}(\hat{p}_{Y_i}, y_i)$ are a sample of $N$ i.i.d. uniforms. Any uniformity test on the realised $u_i$ is therefore a calibration test, with no $K$-way family of correlated sub-tests to correct for; the *rank histogram* for ensembles (Hamill, 2001) is the same idea in discrete form. A quick visual diagnostic comes from the PIT histogram itself (Fig. 9, top row) (Diebold et al, 1998; Gneiting et al, 2007): a U-shape signals over-confidence, a hump signals conservatism, and a slope signals bias in the predictive mean.

Raw $u_i$ is, however, not the most convenient parametrisation for a regression-calibration test, because the coverage view of Section 3 already has a natural coordinate $\lambda$. Folding the PIT around its midpoint aligns the two views exactly:

$$V_i = |2U_i - 1|, \quad v_i = |2 \cdot \text{cdf}(\hat{p}_{Y_i}, y_i) - 1|. \tag{10}$$

For $U \sim \mathcal{U}(0,1)$, $|2U - 1| \sim \mathcal{U}(0,1)$ too, so under calibration $V_i \sim \mathcal{U}(0,1)$ i.i.d. The reason to fold is a one-line identity: the sample $y_i$ lies in the centered prediction interval at level $\lambda$ iff $\text{cdf}(\hat{p}_{Y_i}, y_i) \in [(1-\lambda)/2, (1+\lambda)/2]$ iff $v_i \leq \lambda$, and so

$$\hat{c}(\lambda) = \frac{1}{N} \sum_{i=1}^{N} \mathbb{1}[v_i \leq \lambda] = \hat{F}_v(\lambda). \tag{11}$$

The empirical coverage curve and the empirical CDF of the folded PIT are mathematically the same random object, a direct consequence of the coverage–PIT equivalence of (Gneiting et al, 2007) stated explicitly in the ML calibration context by (Podsztavek et al, 2024) and implemented in the ArviZ library (Säilynoja et al, 2022). The bottom row of Fig. 9 is the empirical coverage curve read as a density over the level axis $\lambda = v$: over-confidence concentrates mass in the right tail (the model spends its probability near $v = 1$ instead of spreading evenly), conservatism concentrates mass near $v = 0$, and bias produces an attenuated asymmetry. The folded view gives a direct visual read of whether the intervals are too narrow or too wide, which is precisely what a coverage practitioner wants.

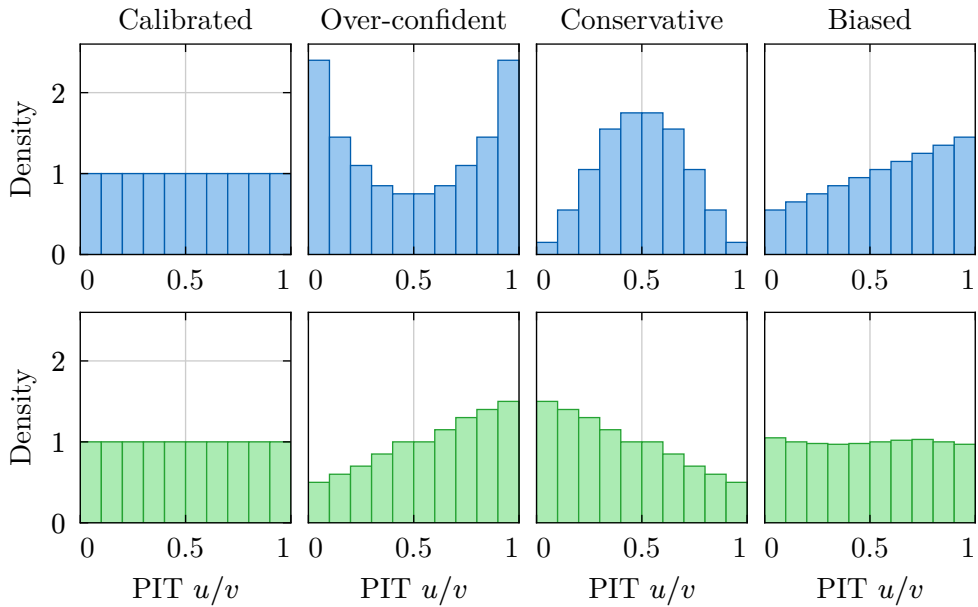

Fig. 9: PIT diagnostics. Top row: raw PIT histograms for four characteristic cases. Bottom row: the corresponding *folded* PIT $v_i = |2u_i - 1|$, which by (11) is a reparametrisation of the empirical coverage curve ($\hat{c}(\lambda) = \hat{F}_v(\lambda)$). Over-confidence appears as a U-shape in the raw histogram and as right-skewed mass in the folded histogram; conservatism as a hump in the raw and left-skewed mass in the folded; bias as a slope in the raw and a milder folded asymmetry. The KS test on the folded PIT (equivalently, on the coverage curve) detects departures in any direction.

The natural joint test against the null $V_i \sim \mathcal{U}(0,1)$ is the two-sided Kolmogorov–Smirnov statistic on the folded PIT,

$$D_N = \sup_{\lambda \in [0,1]} |\hat{F}_v(\lambda) - \lambda|, \tag{12}$$

which by (11) is the same statistic as the KS test on the coverage curve, is distribution-free under $H_0$ by the classical Kolmogorov result (Kolmogorov, 1933; Smirnov, 1939; Shorack and Wellner, 1986), and has asymptotic critical value $c_\alpha = \sqrt{\ln(2/\alpha)/(2N)}$ at size $\alpha$ (a finite-sample-valid upper bound that matches the Kolmogorov tail constant closely). At $N = 365$ this gives $c_\alpha \approx 0.071$; at $N = 100$, $\approx 0.136$. Unlike the coverage Bonferroni, the KS test is a single finite-sample-valid joint test that exploits the exact Brownian-bridge correlation between levels (Section 4.2) rather than assuming worst-case dependence, recovering the statistical power Bonferroni silently gives up.

The recipe generalises beyond the univariate case without any change to the testing machinery. For multivariate Gaussian predictions the PIT is built from the squared Mahalanobis distance (Section 6.1): $U_i = \text{cdf}(\chi_d^2, D_i^2)$ with $D_i^2 = (\boldsymbol{Y}_i - \boldsymbol{\mu}_i)^T \boldsymbol{\Sigma}_i^{-1} (\boldsymbol{Y}_i - \boldsymbol{\mu}_i)$, which is $\chi_d^2$ under calibration and feeds the same KS test after folding $U_i$. The hard requirement is the CDF itself; for predictions that do not admit one (particle clouds, conformal sets), the coverage view of Section 3 remains available and is taken up in Section 6.2 and Section 6.3.

**4.2 The Brownian Bridge between PIT and Coverage**

The KS test of Section 4.1 replaces $K$ Bonferroni-corrected per-level Binomials with a single joint uniformity test on the same random object ((11)). What has not yet been explained is *why* that joint test is strictly more powerful than the Bonferroni-corrected collection: why is Bonferroni giving up power at all?

The answer is that Bonferroni is a worst-case bound over arbitrary joint distributions of the $K$ per-level test statistics (Dunn, 1961), and the actual joint distribution of the coverage counts is extraordinarily far from that worst case. Adjacent coverage tests share almost all of their indicator variables: $y_i$ lies inside $\mathcal{T}_{\lambda_k}(\hat{p}_{Y_i})$ iff $v_i \leq \lambda_k$, so the only indicators that flip between $\lambda_k$ and $\lambda_{k+1}$ are those for samples whose folded PIT lies in the slice $(\lambda_k, \lambda_{k+1}]$. The empirical coverages at adjacent levels are therefore almost perfectly correlated, and Bonferroni's worst-case independence assumption is the opposite of the truth.

A correlation-aware correction can be imagined as a thought experiment. Test the levels sequentially: at $\lambda_1$ unconditionally at size $\alpha$; at $\lambda_2$ condition on $\hat{c}(\lambda_1)$, whose conditional distribution is much tighter than the marginal because adjacent indicators agree almost everywhere; at $\lambda_k$ condition on the prefix $\hat{c}(\lambda_1), ..., \hat{c}(\lambda_{k-1})$. With this running conditioning the joint family-wise error rate is controlled exactly and the per-step thresholds shrink far below Bonferroni's $\tau_p/K$. This is not a practical algorithm (the closed-form conditional Binomial tail is unwieldy), but it shows that the correlation-aware joint test exists and is much tighter than Bonferroni.

In the dense-grid limit $K \to \infty$ the coverage curve $\lambda \mapsto \hat{c}(\lambda) = \hat{F}_{v(\lambda)}$ is *one* random object, the empirical CDF of $N$ i.i.d. uniform variables, not $K$ independent Binomials. A short calculation gives its exact joint covariance: for $s, t \in [0, 1]$,

$$\text{Cov}(\hat{c}(s), \hat{c}(t)) = \frac{\min(s,t) - st}{N}, \tag{13}$$

the classical *Brownian bridge* covariance kernel (Shorack and Wellner, 1986). The marginal variance $\lambda(1-\lambda)/N$ recovers the per-level Binomial variance and the off-diagonals capture the correlation diagnosed above; the sequential conditioning thought experiment is the incremental form of this joint distribution. The joint test that exactly accounts for this covariance is the two-sided Kolmogorov–Smirnov test (12) already introduced in Section 4.1. Coverage and PIT are two windows on the same random object, and the joint test reads the same in both languages; the Bonferroni-vs-KS gap is the price the coverage view pays for keeping its $K$-test phrasing instead of taking the dense limit.

Per-level interpretability is recovered for free by inverting the KS critical value into a uniform two-sided envelope around the diagonal:

$$\text{accept iff} \quad |\hat{c}(\lambda) - \lambda| \leq c_\alpha \quad \text{for all } \lambda \in [0,1], \quad c_\alpha = \sqrt{\ln(2/\alpha)/(2N)}. \tag{14}$$

A practitioner plots $\hat{c}(\lambda)$ against the diagonal exactly as in Section 3, but with a band that knows about the joint correlation, and the envelope is uniformly tighter than per-level Bonferroni across most of the central range of $\lambda$. Against the variance-inflation alternative $Y_i \sim \mathcal{N}(0, (1+\varepsilon)^2)$ with $\varepsilon = 0.10$ at $N = 100$, $K = 9$, $\alpha = 0.05$, the KS test rejects with meaningfully higher probability than Bonferroni, and the gap widens as $\varepsilon$ grows.

### 4.3 Variance Binning for Conditional Calibration

Every metric so far (per-level coverage and the folded-PIT KS test) measures *marginal* calibration: the fraction of samples covered, averaged over inputs. A model can pass marginal calibration while being seriously miscalibrated *conditionally*, over-confident where it reports low uncertainty and under-confident where it reports high uncertainty, with the two errors cancelling in aggregate (Levi et al, 2022; Gneiting and Resin, 2023). For safety this is the worst failure mode: the model looks calibrated overall, yet its uncertainty estimates are unreliable precisely when it claims to be certain.

Variance binning is not a new test, it is a wrapper. Partition the $N$ samples into $K_{\text{bins}}$ disjoint bins by predicted uncertainty (e.g., $\sigma_i$ for univariate predictions or $\det(\boldsymbol{\Sigma}_i)$ for multivariate) and run any of the previous metrics inside each bin. Bins are typically defined by quantiles of $\sigma_i$ for roughly equal sample sizes per bin. The strongest property a binned check targets is

$$\forall \sigma_0: \quad \mathbb{E}\left[(\mu_i - Y_i)^2 \mid \sigma_i = \sigma_0\right] = \sigma_0^2, \tag{15}$$

i.e., the predicted variance equals the true conditional variance at every reported uncertainty level.

Unlike per-level coverage tests, different variance bins use *disjoint* samples, so the tests across bins are genuinely independent and Bonferroni across bins is tight: this is the one place in the paper where the worst-case dependence assumption matches the actual dependence. The cost of binning is per-bin sample size: roughly $N/K_{\text{bins}}$ samples per bin, with $K_{\text{bins}} \in \{3, 5, 10\}$ a typical compromise. The hypothesis refinements of Section 5 (one-sided and tolerance) apply directly within each bin.

### 4.4 Trade-offs: Coverage, KS, and Binning

The three metric families trade power, interpretability, and conditional sensitivity differently (Table 1).

The folded KS band *replaces* Bonferroni-corrected per-level coverage as the metric workhorse, recovering per-level interpretability via the uniform envelope of (14) while picking up substantial statistical power. Variance binning is *orthogonal* and addresses conditional miscalibration; the two compose freely.

## 5 The Safety-Critical Hypothesis: One-Sided Tests and Tolerance Bands

The hypothesis slot of Section 3 is a classical two-sided Binomial test: it rejects whenever the observed coverage is either too low or too high. That default carries two drawbacks for safety-critical deployment, and this section addresses each with an independent refinement. Neither refinement is novel in isolation, but taken together they turn the basic pipeline of Section 3 into something an engineer can actually write into a safety case, and that assembly is the paper's safety-critical contribution.

The first drawback is directional. Safety is asymmetric: a model whose intervals are too narrow understates uncertainty and exposes downstream planners to unaccounted-for risk, while a model whose intervals are too wide merely sacrifices sharpness. A test that rejects both equally is the wrong shape for a safety case. Section 5.1 replaces the two-sided test with a one-sided variant that accepts conservatism as non-dangerous, introduces the *folded* PIT as the same directional reduction applied to the KS test, and closes with a practical observation: when the one-sided test rejects, the cleanest operational remedy is to inflate the predicted variances.

|  | **Coverage + Bonferroni** | **Folded KS band** | **Variance binning** |
|---|---|---|---|
| Joint correlation | Ignored | Exact (Brownian bridge) | Per-bin (independent) |
| Interpretability | High (per-level %) | High (uniform band on coverage) | Per-bin coverage |
| Conditional failures | Hidden | Hidden | Surfaced |
| Requires CDF | No | Yes (PIT) | Inherits inner metric |
| Cost | $O(K)$ | $O(N \log N)$ | $K_{\text{bins}} \times$ inner |

Table 1: Comparison of the metric families considered in this paper. Coverage with Bonferroni is the simplest; the folded KS band gives the same per-level interpretability via a uniform envelope on the coverage curve and exploits the joint Brownian-bridge correlation structure exactly; variance binning is the only one of the three that can surface conditional miscalibration. The hypothesis-slot refinements of Section 5 (one-sided and tolerance) apply to all three.

The second drawback is asymptotic. With enough data, any approximate model is eventually rejected, no matter how negligible the miscalibration (Berkson, 1938; Meehl, 1967): a model with a tiny variance inflation $\varepsilon > 0$,

$$Y_i \sim \mathcal{N}\big(0, (1+\varepsilon)^2\big), \quad H_0 : Y_i \sim \mathcal{N}(0,1), \tag{16}$$

gives a standardised statistic $Z_n \approx \varepsilon\sqrt{2n}$, so $p_n \to 0$ regardless of how small $\varepsilon$ is, and a model with 0.1% miscalibration is eventually rejected even though it is perfectly adequate in practice. Fig. 10 illustrates the pathology: the two-sided acceptance band at a single coverage level shrinks with $N$ and eventually excludes the true coverage of a mildly miscalibrated model. Section 5.2 replaces the exact null with a *tolerance* region, widening the acceptance set by an engineering-specified $\varepsilon_{\text{tol}}$ via the logic of equivalence testing (Schuirmann, 1987; Lakens, 2017).

The two refinements are independent and compose freely. They apply to every metric of Section 4 (per-level coverage, folded KS, variance binning), to every model-and-data extension of Section 6, and to the online testing of Section 7.

### 5.1 One-Sided Hypotheses

The simplest route to a safety-critical hypothesis is to keep the per-level Binomial of Section 3 but reject only when the coverage count is too low:

$$\text{Reject } H_0 \text{ at level } \tau_p \quad \text{iff} \quad c_{\lambda_k} < \text{qtl}\big(\text{Binomial}(N, \lambda_k), \tau_p\big). \tag{17}$$

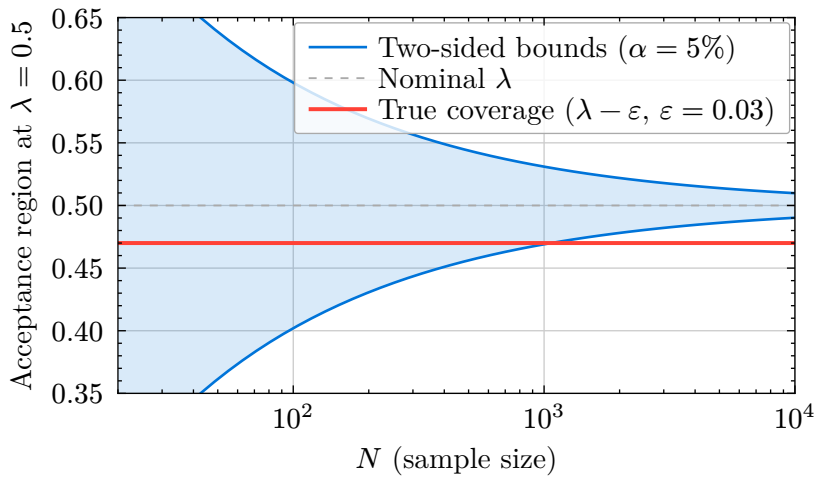

Fig. 10: The large-$N$ problem: the two-sided acceptance band (blue) shrinks as $N$ grows, eventually excluding a model with only $\varepsilon = 0.03$ miscalibration (red line). Tolerance bands (Section 5.2) widen the acceptance region to prevent this.

A conservative model with wide intervals produces high counts and is never rejected, which is exactly the desired behaviour; only the direction that actually endangers downstream use is treated as evidence against the model.

The same directional move applies to the KS test of Section 4.1, via the folded PIT $v_i = |2u_i - 1|$ that already underwrites the two-sided KS test and the Brownian-bridge bridge of Section 4.2. Over-confidence concentrates mass in the right tail of the folded distribution (cf. Fig. 9), while conservatism and bias push mass toward the left. The one-sided analogue of (12) therefore tests only the right-tail departure from uniformity,

$$D_N^+ = \sup_{\lambda \in [0,1]} \left( \lambda - \hat{F}_v(\lambda) \right), \tag{18}$$

and its size-$\alpha$ critical value is the classical one-sided Kolmogorov–Smirnov constant

$$c_\alpha^+ = \sqrt{\ln(1/\alpha)/(2N)} \tag{19}$$

(Smirnov, 1939; Shorack and Wellner, 1986). At $N = 365$ this gives $c_\alpha^+ \approx 0.064$; at $N = 100$, $\approx 0.122$, slightly sharper than the two-sided thresholds of Section 4.1 because half the statistic's range is discarded by construction. Per-level interpretability returns via a one-sided lower envelope of the coverage curve, $\hat{c}(\lambda) \geq \lambda - c_\alpha^+$ for all $\lambda$, which is the one-sided variant of (14).

The folded PIT is in this reading the safety-critical reduction of the full PIT: systematic bias in the predictive mean leaves $u_i$ slanted but does not push folded values into the right tail, so the one-sided folded KS test is effectively *blind to bias*. A model that is "wrong on average but safe in dispersion" passes the test, which is precisely the desired behaviour when safety certification does not care about mean error as long as the uncertainty envelope is honest. This is also the step where the folding transformation carries a contribution of its own: the idea of $|x - \text{median}(x)|$ as a scale diagnostic appears in MCMC convergence (Vehtari et al, 2021), folded rank statistics show up in simulation-based calibration (Modrák

et al, 2023), and the ArviZ library implements the same reduction for coverage visualisation (Säilynoja et al, 2022), but to our knowledge its use as a one-sided KS statistic for safety-critical calibration is new.

The natural operational fix, when the one-sided test rejects, is to inflate predicted variances uniformly until the test passes. In a safety-critical setting this is always directionally correct, because inflating variance forces downstream planners to act more conservatively, which is by construction what we want. The cost is sharpness: wider intervals mean less confident decisions and typically worse task performance, exactly the tradeoff the calibration check is designed to expose. Performance should only be chased once the safety margin is in place, and a calibration-driven variance inflation is the cleanest way to enforce that ordering.

**5.2 Tolerance Bands**

The second refinement replaces the exact null with a *tolerance* null that admits a small, engineering-specified $\varepsilon_{\mathrm{tol}}$ worth of miscalibration as acceptable. Equivalence testing has been applied to deterministic model validation in ecology (Robinson and Froese, 2004) and to expert-elicitation calibration with small samples (Dharmarathne et al, 2022), and the well-documented large-$N$ pathology of standard calibration tests (Van Calster et al, 2019; Austin and Steyerberg, 2020) has been studied from a property-testing angle (Błasiok et al, 2024); to our knowledge, it has not previously been applied to probabilistic regression calibration, where it composes cleanly with every other slot of the pipeline.

For the per-level Binomial coverage test, tolerance widens the acceptance region by testing against *shifted* Binomials rather than the exact null:

$$\begin{aligned}\text{reject iff} \quad & c(\lambda) < \mathrm{qtl}(\mathrm{Binomial}(N, \lambda - \varepsilon_{\mathrm{tol}}), \alpha/2) \\ & \text{or } c(\lambda) > \mathrm{qtl}(\mathrm{Binomial}(N, \lambda + \varepsilon_{\mathrm{tol}}), 1 - \alpha/2).\end{aligned} \quad (20)$$

Typical values of $\varepsilon_{\mathrm{tol}}$ range from 0.01 to 0.05 and encode a pragmatic engineering judgment of how much miscalibration is acceptable. Composed with the one-sided refinement of Section 5.1, the rule becomes "accept whenever $\hat{c}(\lambda) \geq \lambda - \varepsilon_{\mathrm{tol}}$ for all $\lambda \in \Lambda$": conservatism is always acceptable, small over-confidence is tolerated, and only miscalibration that exceeds the engineering budget is rejected.

The same shift extends to the folded KS band of (14). The envelope absorbs a tolerance simply by translating the diagonal,

$$\text{accept iff} \quad \hat{c}(\lambda) \geq \lambda - c_\alpha^+ - \varepsilon_{\mathrm{tol}} \quad \forall \lambda \in [0,1], \quad (21)$$

shifting the null from "exactly calibrated" to "true coverage at least $\lambda - \varepsilon_{\mathrm{tol}}$ at every level" and preserving the KS guarantee (the rejection probability under the shifted null is still at most $\alpha$). The result is a one-sided equivalence test (Schuirmann, 1987) with a clean engineering reading: "accept any model whose true coverage is at most $\varepsilon_{\mathrm{tol}}$ short of nominal". This phrasing is analogous to the integrity-risk budget in GNSS receiver certification (Blanch et al, 2015) and produces exactly the kind of specification that fits directly into a safety case, in contrast to a raw p-value which requires the reader to interpret statistical significance. For applications where any deviation from uniformity (including bias) must be detected, the two-sided KS test of Section 4.1 remains available without tolerance; for safety-critical use, the one-sided folded KS band with an explicit tolerance is the natural tool.

## 6 Refining the Model & Data: Multivariate, Non-Gaussian, and Particles

The metric, hypothesis, and testing machinery built so far generalises essentially unchanged when the univariate Gaussian is replaced by multivariate Gaussian (Section 6.1), other parametric families (Section 6.2), or particle-based non-parametric predictions (Section 6.3). The inner slots stay fixed; only the data model swaps.

### 6.1 Multivariate Gaussian

For a multivariate Gaussian $\mathcal{N}(\boldsymbol{\mu}, \boldsymbol{\Sigma})$ with $\boldsymbol{\mu} \in \mathbb{R}^d$, points have no natural order and quantile-based intervals do not generalise directly (Gneiting et al, 2008). Instead, use the squared *Mahalanobis distance* (Mahalanobis, 1936):

$$D_i^2 = (\boldsymbol{y}_i - \boldsymbol{\mu}_i)^T \boldsymbol{\Sigma}_i^{-1} (\boldsymbol{y}_i - \boldsymbol{\mu}_i). \tag{22}$$

Under calibration $D_i^2 \sim \chi_d^2$, and the coverage set at level $\lambda$ is the ellipsoid

$$\mathcal{T}_\lambda(\mathcal{N}(\boldsymbol{\mu}, \boldsymbol{\Sigma})) = \{\boldsymbol{y} : (\boldsymbol{y} - \boldsymbol{\mu})^T \boldsymbol{\Sigma}^{-1} (\boldsymbol{y} - \boldsymbol{\mu}) \leq \mathrm{qtl}(\chi_d^2, \lambda)\}. \tag{23}$$

The rest of the pipeline applies unchanged with ellipsoidal sets replacing scalar intervals. The scalar PIT $u_i = \mathrm{cdf}_{D_i^2}(\chi_d^2)$ maps each prediction–observation pair to a uniform value and feeds the KS / folded-KS machinery directly. Variance binning bins by $\det(\boldsymbol{\Sigma}_i)$ (or its logarithm) to probe conditional calibration across uncertainty magnitudes.

### 6.2 Non-Gaussian Parametric and Prediction-Set Methods

The coverage pipeline applies to *any* system that produces prediction sets; the PIT is not required.

**Parametric families with tractable quantiles.** For distributions with closed-form quantile functions (Student-$t$, Exponential, Gamma) the centered intervals of (2) are defined directly; for mixtures and similar, numerical CDF inversion suffices. The univariate PIT $U_i = \mathrm{cdf}(\hat{p}_{Y_i}, Y_i)$ is also available, so coverage and folded-KS pipelines both apply. The clean multivariate reduction of Section 6.1 via Mahalanobis $\chi_d^2$ is, however, Gaussian-specific: for general multivariate non-Gaussian parametric models no analogous scalar reduction exists, and coverage-based methods become the primary tool.

**Prediction sets without densities.** Conformal methods (Vovk et al, 2005; Romano et al, 2019) output a set $\mathcal{T}_\lambda(x_i)$ at each level with a marginal coverage guarantee, but no underlying density. The coverage pipeline (Section 3) handles this directly: check $y_i \in \mathcal{T}_\lambda(x_i)$ and proceed with Binomial testing. The folded-KS test is unavailable (no PIT to fold), but per-level Bonferroni and variance binning still apply. This makes coverage-based calibration checking the natural validation tool for set-valued predictors. The only requirement is the ability to evaluate the indicator $\mathbb{1}[y \in \mathcal{T}_\lambda(\hat{p})]$.

### 6.3 Non-Parametric Predictions

Particle-based predictions $\{(w^j, \hat{\boldsymbol{y}}^j)\}_{j=1}^M$, e.g., from a particle filter, have no closed-form CDF or quantile function, so the PIT is simply undefined and the folded-KS route of

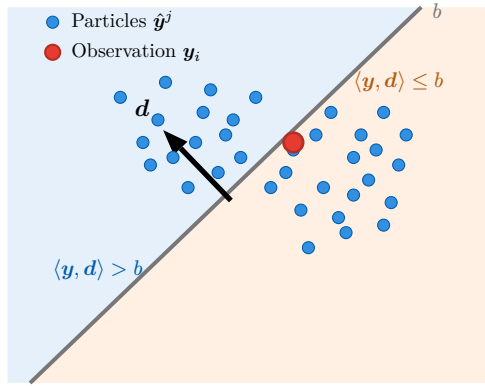

Fig. 11: **Half-plane method.** A random direction $d$ and threshold $b$ partition the space into two half-planes. The predicted particles $\hat{y}^j$ (blue) and observed outcome $y_i$ (red) are classified by which side they fall on. Under calibration, the weighted particle fraction above the cut should match the observation's indicator.

Section 4.1 is unavailable. The coverage view of Section 3, which needs only the indicator $\mathbb{1}[y \in \mathcal{T}_\lambda(\hat{p})]$ and not an underlying density, remains applicable, and the *sliced Wasserstein* idea (Rabin et al, 2012; Bonneel et al, 2015) generalises it to the multivariate case: reduce the multivariate comparison to one-dimensional checks along random projections, each of which is a scalar coverage test. Dimension-reduction strategies for multivariate calibration assessment have been explored in the ensemble-forecasting literature, notably the multivariate rank histograms and pre-rank functions of (Gneiting et al, 2008) and the energy-score framework of (Allen et al, 2024); the construction below specialises the idea to weighted particle clouds in a way that preserves the Bernoulli/Binomial coverage structure, so the entire downstream pipeline (one-sided hypotheses, Bonferroni or KS, tolerance bands, e-values) carries over without modification.

For a random direction $d \in \mathbb{R}^p$ and a threshold $b$, define the *half-plane indicator* $h_{d,b}(y) = \mathbb{1}[\langle y, d \rangle > b]$ (Fig. 11). For an observation $y_i$ and particle set $\hat{P}_i = \{(w^j, \hat{y}^j)\}$, compare the predicted and observed proportions above the cut:

$$\Delta_{d,b} = \sum_j w^j \cdot h_{d,b}(\hat{y}^j) - h_{d,b}(y_i). \qquad (24)$$

Under calibration $\mathbb{E}[\Delta_{d,b}] = 0$ for every $(d, b)$.

The identity holds for *every* $(d, b)$, but in practice only a finite collection can be tested and the choice is what governs statistical power. A natural construction samples a random direction $d \sim \text{Uniform}(S^{p-1})$, projects the predicted particles onto that direction to obtain a one-dimensional predicted distribution, and takes the thresholds as its weighted empirical quantiles at a small grid of levels $(0.05, 0.10, ..., 0.95)$. With $b$ chosen this way the half-plane indicator along $d$ is Bernoulli with known mean by construction, and sweeping $b$ over the predicted quantiles reproduces the per-level coverage curve of Section 3.2 exactly: the multivariate problem reduces to the one-dimensional coverage check of Section 3 along each sampled direction, and the downstream pipeline (one-sided hypotheses, Bonferroni or KS, tolerance bands, e-values) carries over unchanged.

**Connection to coverage.** In one dimension ($p = 1$, $d = 1$) this reduces to comparing the particle ECDF at $b$ against $\mathbb{1}[y_i > b]$, and sweeping $b$ traces out the survival function of the predicted distribution: it is the coverage check of Section 3. For the multivariate version we sample a collection $\{(d_\ell, b_\ell)\}_{\ell=1}^{L}$ of directions and cuts *before looking at the data*, and freeze them. Conditional on $(d_\ell, b_\ell)$, the indicator $h_{d_\ell, b_\ell}(y_i)$ is Bernoulli with null mean $\sum_j w^j h_{d_\ell, b_\ell}(\hat{y}^j)$, and the count over $N$ samples is Binomial just like the coverage counts of Section 3. The entire downstream machinery (one-sided hypotheses, Bonferroni or folded KS, tolerance bands) carries over to the half-plane counts. Bonferroni across the $L$ directions is appropriate because the directions are sampled independently of the data.

The random directions also recover Wasserstein in expectation,

$$W_1(\hat{P}, \delta_y) = \mathbb{E}_{d \sim S^{p-1}} \left[ W_1(\text{proj}_d \hat{P}, \text{proj}_d \delta_y) \right], \tag{25}$$

so the sliced approach inherits the well-known Wasserstein guarantees in the limit $L \to \infty$.

**Choosing $L$ with the state-space dimension.** A finite collection of $L$ random directions probes only an $L$-dimensional slice of the $(p-1)$-sphere, so miscalibration concentrated in a narrow band of directions (e.g., a single under-estimated eigendirection of the true covariance) becomes progressively easier to miss as the ambient dimension $p$ grows. The Monte-Carlo rate at which the sliced-Wasserstein average approaches its population expectation is $O(L^{-\frac{1}{2}})$ (Bonneel et al, 2015), with constants that grow at most polynomially in $p$ for distributions with bounded moments, so a *linear* scaling $L \propto p$ (or $L \propto p \log p$ if a large safety margin is desired) is sufficient to keep the directional approximation error bounded as the dimension grows. In practice we recommend the scaling $L = c \cdot p$ with $c \in \{5, 10, 20\}$, treating $c$ as a tuning knob on a held-out validation fold and increasing it until the test decision stabilises. In the $p = 2$ robot experiment of Section 8.2 we use $L = 20$, well above the asymptotic requirement; for full $p = 6$ rigid-body pose one would move to $L \in [30, 60]$, and for higher-dimensional states (e.g., joint-space configurations in $p = 10$–$20$) a deliberate sweep in $L$ is the honest way to report dimensional sensitivity. The Bonferroni factor $\alpha/(LK)$ grows only logarithmically in $L$, so the power cost of a generous $L$ is mild.

The method is necessarily less sharp than its parametric counterparts, but it handles arbitrary predictive distributions without distributional assumptions while preserving the coverage-based testing framework.

## 7 Refining the Testing: E-values and Online Monitoring

The pipeline of Section 3 used Bonferroni-corrected p-values: the offline batch paradigm of computing one number on a fixed validation set and deciding once. Safety-critical deployment often asks for something stronger, a test that can be evaluated *continuously* as predictions arrive, with an alarm that fires the moment evidence crosses a threshold. P-values fail this on three counts: combining them across coverage levels requires conservative corrections, the sample size $N$ must be fixed in advance (so peeking invalidates the test), and continuous monitoring is incompatible with the p-value definition.

*E-values* address all three (Shafer, 2021; Vovk and Wang, 2021; Ramdas et al, 2023; Grünwald et al, 2024). An e-value $E$ is a non-negative random variable with $\mathbb{E}[E \mid H_0] \leq 1$, and we reject $H_0$ when $E > 1/\alpha$.

| Scenario | P-value | E-value |
| --- | --- | --- |
| Single test, fixed $N$ | Appropriate | Appropriate |
| Multiple coverage levels | Requires correction | Average directly |
| Online monitoring | Invalid | Anytime-valid |
| Combine across experiments | Complex | Multiply or average |

Table 2: Comparison of p-values and e-values for different use cases.

**Three properties make e-values fit the runtime setting.** (i) The average $\overline{E} = K^{-1} \sum_k E_k$ of any (possibly dependent) e-values is a valid e-value (Vovk and Wang, 2021): there is no multiple-testing correction. (ii) The running product $E_t$ may be inspected at any stopping time, with Type-I error control preserved by Ville's inequality (Ville, 1939). (iii) These two together yield genuine online safety monitoring (Howard et al, 2021; Arnold et al, 2023), which is precisely the runtime assurance primitive deployed systems need. For sequential calibration testing specifically, (Henzi and Ziegel, 2022) build e-value tests for proper scoring rules and (Casgrain et al, 2024) extend the construction to general elicitable functionals via supermartingales; the likelihood-ratio test martingale of (26) below is a textbook instance of their machinery applied to a Binomial coverage model. Our contribution in this section is therefore not the e-value construction itself but its integration: the testing slot composes with every metric variant (coverage, folded KS, variance binning) and every hypothesis variant (one-sided, tolerance) developed in Section 4 and Section 5, and the drift-sweep monitor of Section 8.2 demonstrates the resulting safety primitive end-to-end. From the practitioner's perspective, the key observation is that swapping the testing slot from p-values to e-values requires no changes to the upstream metric or hypothesis choices.

### 7.1 Construction for Coverage

For coverage-based checks with a Binomial model, the likelihood-ratio e-value is

$$E = \left(\frac{p_{\text{alt}}}{p_{\text{null}}}\right)^k \left(\frac{1 - p_{\text{alt}}}{1 - p_{\text{null}}}\right)^{N-k}, \qquad (26)$$

with $k$ the observed count, $p_{\text{null}} = \lambda$, and $p_{\text{alt}} < \lambda$ the over-confidence alternative. For sequential use, the running product

$$E_t = \prod_{i=1}^{t} e_i \qquad (27)$$

is a *test martingale* that may be evaluated at any stopping time. The threshold $1/\alpha$ becomes the engineering spec: the runtime monitor raises an alarm when cumulative evidence of overconfidence crosses it, regardless of when the evidence arrives or how many predictions have been seen.

|  | Weather forecasting | Robot localization |
|---|---|---|
| **Model & Data** | Univariate Gaussian | Particle cloud ($M = 500$) |
| **Metric** | PIT uniformity | Half-plane coverage |
| **Hypothesis** | Two-sided | One-sided (safety-critical) |
| **Testing** | Offline KS $p$-value | Online likelihood-ratio e-value |

Table 3: **Two example recipes drawn from the framework.** Weather uses an offline distributional test suited to retrospective validation; the robot localizer uses an online safety-critical monitor suited to runtime assurance. Every slot differs between the two recipes, and the framework supports many more combinations (Appendix A).

## 8 Experiments

The point of the four-slot framework is composability: any one slot can be swapped without re-deriving the others, and the same grammar yields recipes that look operationally different on the same data. We demonstrate this on two safety-critical problems that sit at opposite corners of the design space, and instantiate one recipe on each. Table 3 summarises the two recipes side by side; every slot differs. The framework supports many more combinations, and Appendix A walks through four slot-swap ablations on the same datasets to show how each slot can be varied independently.

### 8.1 Weather Temperature Prediction

We predict tomorrow's daily temperature high at a fixed weather station, a standard testbed for probabilistic forecasting (Gneiting et al, 2005; Thorarinsdottir and Gneiting, 2010). A sliding-window autoregression on the preceding $W = 30$ days produces

$$\hat{p}_{Y_t}(x_t) := \mathcal{N}(\hat{\mu}_t, \hat{\sigma}_t^2), \tag{28}$$

with $\hat{\mu}_t$ the linear-regression prediction and $\hat{\sigma}_t^2$ the in-window residual variance. The setting is well-suited to a Gaussian assumption, provides $N = 365$ daily predictions (year 2016), and is a natural offline validation target: we retrospectively ask whether a year of forecasts was distributionally faithful before signing off on the model. The AR forecaster is unbiased on average ($\hat{\mu} = 16.9°$ C matches the empirical mean to a tenth of a degree) and its mean predicted standard deviation is $\hat{\sigma} = 3.2°$ C.

The recipe fixes the metric slot to PIT uniformity (Section 4.1) and the testing slot to an offline Kolmogorov–Smirnov $p$-value on the PIT ECDF. The closed-form Gaussian CDF makes this the natural metric on parametric forecasts: $K$ Bonferroni-corrected coverage Binomials are replaced by a single uniformity test that respects the Brownian-bridge correlation structure (Section 4.2). Within this recipe, the hypothesis slot can be read in two operationally different ways on the same PIT sample, and we report both in parallel to show what each one picks up: a two-sided test on the raw PIT, which detects *any* departure from uniformity (Fig. 13), and a one-sided test on the folded PIT $v = |2u - 1|$ with a small engineering tolerance $\varepsilon_{\text{tol}} = 0.02$, which only rejects if the forecast is *over-confident* by more than the tolerance (Fig. 14).

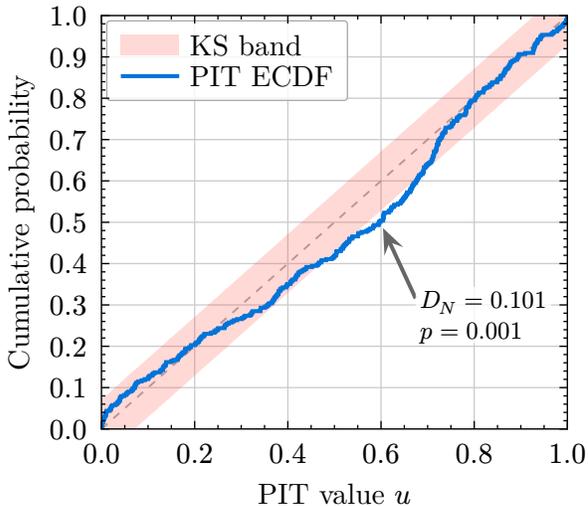

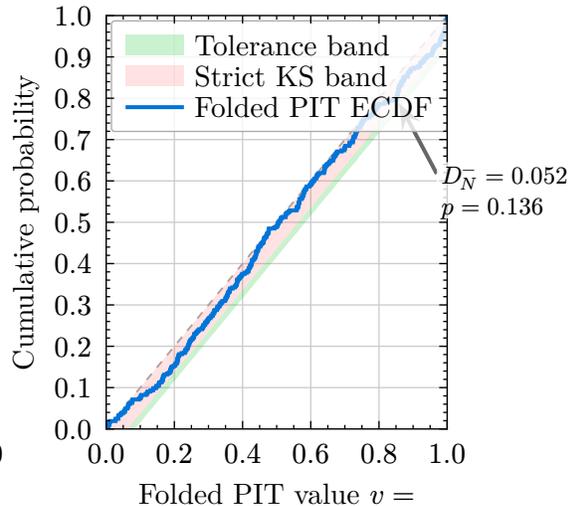

Fig. 13: **Two-sided KS on the raw PIT.** The ECDF departs from the uniform diagonal and exits the two-sided KS band; the test rejects at $p = 0.001$. The deviation is concentrated on one side of the diagonal, diagnostic of a systematic *location bias* in the forecaster's mean rather than a dispersion failure.

Fig. 14: **One-sided KS with tolerance on the folded PIT.** The folded ECDF $F_{N(v)}$ stays inside both the strict KS band (red) and the tolerance-extended band (green, $\varepsilon_{\text{tol}} = 0.02$); the safety-critical test passes. Folding removes the location-bias signal and keeps only the over-confidence signal, which here is within the engineering tolerance.

The two readings answer different questions on the same data. The two-sided KS asks "is the forecast *distributionally* faithful?", and the answer is no: the ECDF sits coherently on one side of the diagonal because the AR forecaster's conditional mean carries a small residual bias. The one-sided folded KS asks the narrower operational question "are the stated prediction intervals wide enough?", and the answer is yes, within the 2% tolerance. The bias-blindness of the folded reading is deliberate (Section 5.1): a forecasting scientist concerned with unbiasedness would flag the model for retraining, a safety engineer concerned with over-confidence would cite the folded reading as evidence of fitness for deployment, and both conclusions are legitimate. The slot swaps in Appendix A revisit the same data from additional angles.

**8.2 Robot Localization via Particle Filter**

A robot in a 2D environment with known beacon positions executes a random walk with a small systematic drift and observes noisy range measurements. A particle filter ($M = 500$ particles) tracks it with a motion model that *does not account for the drift*, a realistic source of model mismatch in the spirit of the EKF-SLAM consistency issues of (Bailey et al, 2006) and (Huang et al, 2010).

The filter output is a particle cloud per time step, not a closed-form density, and the setting is sequential: we want a calibration monitor that accumulates evidence as the robot moves and can raise an alarm the instant enough evidence has arrived. This is the polar opposite of the weather recipe on every slot. The data slot becomes non-parametric (no CDF to feed into a PIT), so the metric slot has to operate on particles directly; the failure mode we worry about is over-confidence along the drift direction, so the hypothesis slot becomes one-sided; and the testing slot has to be anytime-valid so that the monitor's decision is licensed at every step, not only after a fixed horizon.

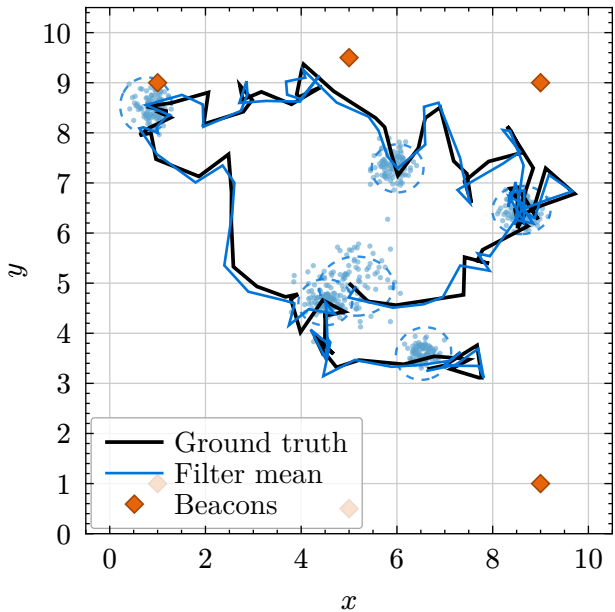

Fig. 15: **Particle filter trajectory.** The filter mean (blue) tracks the ground truth (black) with a small lag due to unmodeled drift in a 2D beacon environment. Particle clouds (light blue dots) are shown at selected time steps.

Recipe 2 spells out the inner loop: at each step, draw a fresh random direction on the unit circle, project the particle cloud and the ground truth onto it, form a level-$\lambda$ weighted empirical interval from the projected particles, and multiply a running test martingale by the per-step likelihood-ratio e-value of the resulting coverage indicator. The construction composes the half-plane metric of Section 6.3 with the sequential e-value of Section 7, cleanly swapping two slots relative to the pedagogical recipe of Recipe 1. Because the direction is sampled independently of the observation, the per-step indicator is Bernoulli ($\lambda$) marginally under calibration, so the test martingale $\mathbb{E}[e_t \mid \mathcal{F}_{t-1}] \leq 1$ holds direction-by-direction and the alarm at the first crossing of $1/\alpha$ is anytime-valid by Ville's inequality.

**Drift sweep.** We run the filter under three drift magnitudes ($0.5 \times$, $1 \times$, $2 \times$ baseline) with 20 seeds each. Fig. 16 shows the median running e-value at $\lambda = 0.9$ with the 5%–95% envelope across seeds. Under benign drift ($0.5 \times$) no seed crosses the rejection threshold $1/\alpha = 20$ within $T = 500$ steps, matching the nominal false-positive control. At baseline drift 3/20 seeds cross, with median first-crossing time $t \approx 330$. At high drift ($2 \times$), 9/20 seeds cross, with median first crossing $t \approx 143$: the monitor fires earlier and more often as the model mismatch grows, which is exactly the runtime-assurance behaviour a safety case needs. Because the per-step direction is randomised, no seed is tuned to the drift, and the signal the monitor picks up is an *average* over the unit circle; adversarial-direction variants that project along a worst-case heading would fire even earlier, at the cost of requiring prior knowledge of the drift.

---

1 **Input:** particle stream $\{(w_t^j, \hat{y}_t^j)\}_{j=1}^M$ and observations $\{y_t\}$ for $t = 1, 2, ...$; coverage level $\lambda$; alternative $p_{\text{alt}} < \lambda$; significance $\alpha$
2 $E_0 \leftarrow 1$
3 **for** $t = 1, 2, ...$ **do**
4     $\theta \sim \mathcal{U}(0, \pi), \quad d \leftarrow (\cos\theta, \sin\theta)$    ▷ fresh direction
5     $z^j \leftarrow \langle \hat{y}_t^j, d \rangle$ for each $j$, $\quad z \leftarrow \langle y_t, d \rangle$    ▷ project particles and observation
6     lo $\leftarrow$ weighted quantile of $\{(w_t^j, z^j)\}$ at $(1 - \lambda)/2$
7     hi $\leftarrow$ weighted quantile of $\{(w_t^j, z^j)\}$ at $(1 + \lambda)/2$
8     $k_t \leftarrow \mathbb{1}[\text{lo} \leq z \leq \text{hi}]$    ▷ coverage indicator, Bernoulli($\lambda$) under $H_0$
9     $e_t \leftarrow (p_{\text{alt}}/\lambda)^{k_t}((1 - p_{\text{alt}})/(1 - \lambda))^{1 - k_t}$    ▷ likelihood-ratio, (26)
10     $E_t \leftarrow E_{t-1} \cdot e_t$    ▷ test martingale
11     **if** $E_t > 1/\alpha$ **then** raise alarm
12 **end**

---

Recipe 2: **Half-plane e-value monitor for particle-filter localization.** A fresh direction on the unit circle is drawn at each step independently of the observation, the particle cloud and the ground truth are projected onto it, and the resulting one-sided coverage indicator at level $\lambda$ feeds a running likelihood-ratio test martingale ((26)). Because the direction is independent of the data, the per-step indicator is Bernoulli($\lambda$) under $H_0$ regardless of direction, so the martingale is valid and the first-crossing alarm is anytime-valid by Ville's inequality. This is the single recipe used on the robot localizer of Section 8.2; relative to the pedagogical Recipe 1 it swaps every slot (particles / half-plane / one-sided / e-value).

**Temporal dependence.** The $T = 500$ samples are consecutive filter outputs, not an i.i.d. draw: adjacent coverage indicators share most of the underlying belief state. The offline KS and Binomial tests of Section 3 and Section 4 would rest on a strictly violated assumption here, and this is the essential reason the robot recipe uses a sequential monitor instead. Test martingales require only $\mathbb{E}[e_t \mid \mathcal{F}_{t-1}] \leq 1$ under $H_0$, which is implied by the conditional calibration condition $P(y_t \in [\text{lo}_t, \text{hi}_t] \mid \mathcal{F}_{t-1}) = \lambda$ (Gneiting et al, 2007), a strictly stronger assumption than marginal calibration that the likelihood-ratio construction (26) exploits directly. The drift sweep of Fig. 16 is therefore valid as reported, without any thinning or block-bootstrap corrections.

**Safety-case interpretation.** A certifier would cite the half-plane e-value monitor as the runtime mechanism for detecting calibration drift in deployment, with the drift sweep as the sensitivity evidence that monitor response scales predictably with the magnitude of model mismatch. The fresh-direction construction means the monitor never assumes a privileged direction of model error, which is appropriate when the deployment environment may induce unanticipated forms of drift. Appendix A revisits this experiment with a fitted Gaussian in place of the particle cloud, showing how the data-model slot can be swapped without touching the rest of the pipeline.

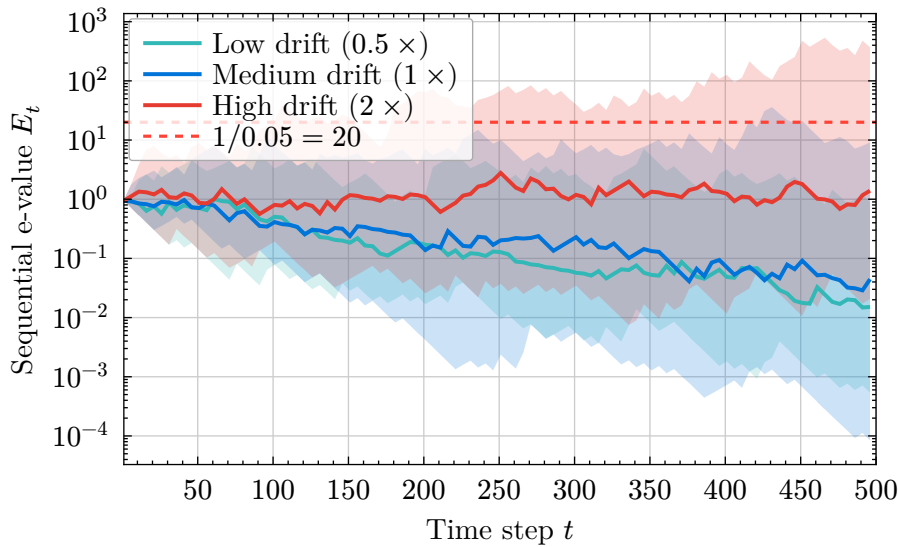

Fig. 16: **Half-plane e-value monitor under three drift magnitudes.** Solid lines: median running e-value at $\lambda = 0.9$ across 20 seeds per drift setting. Shaded bands: 5%–95% envelope. Dashed: rejection threshold $1/\alpha = 20$. Under benign drift (0.5 ×) no seed crosses; at baseline drift 3/20 cross; at high drift 9/20 cross, with median first crossing $t \approx 143$.

## 9 Discussion and Related Work

Our subject is *testing* whether a prediction system is calibrated, as opposed to *achieving* calibration; the two are complementary, and a deployment pipeline would typically run a test first and apply recalibration only on failure. The introduction argued that the regression calibration literature fragments along four largely independent design axes and that these axes have rarely been treated as co-equal slots of a single grammar. We now position the framework against the specific threads it draws on.

**Calibration vs. recalibration.** A large body of work adjusts poorly-calibrated predictions until they pass a calibration check: Platt and temperature scaling (Guo et al, 2017), histogram binning (Naeini et al, 2015), and isotonic regression in classification; isotonic recalibration on the coverage curve for regression (Kuleshov et al, 2018), with extensions by (Song et al, 2019) and (Chung et al, 2021). These methods are orthogonal to ours: they produce a better-calibrated model, we produce a certificate for a given model.

**Proper scoring rules, the coverage–PIT paradigm split, and conformal prediction.** Probabilistic-forecast evaluation rests on *proper scoring rules* (Brier, 1950; Good, 1952; Matheson and Winkler, 1976; Gneiting and Raftery, 2007), which incentivise honest reporting and admit a calibration / sharpness decomposition (Bröcker, 2009) within the (Gneiting et al, 2007; Gneiting and Katzfuss, 2014) paradigm. The ML and conformal communities then evaluate calibration via *coverage* of prediction intervals (Christoffersen, 1998; Kuleshov et al, 2018), while the forecast-verification and Bayesian-computation communities work with *PIT* and rank statistics (Rosenblatt, 1952; Diebold et al, 1998; Hamill, 2001; Gneiting et al, 2008); the two views are formally equivalent (Gneiting et al, 2007) but the equivalence is rarely made explicit in a way that helps a practitioner choose between them, with recent work beginning to bridge this gap (Säilynoja et al, 2022; Podsztavek et al, 2024). Conformal prediction (Vovk et al, 2005; Shafer and Vovk, 2008; Romano et al, 2019; Barber et al, 2021; Barber et al, 2023; 2023) is a *forecasting* method rather than a checking method; because its outputs are sets without underlying densities, the coverage half of our pipeline applies directly (the PIT half does not), and can serve to empirically verify the finite-sample coverage guarantee under deployment and distribution shift.

**PIT folding, equivalence testing, and sequential tests.** The folding transformation that we use to isolate overconfidence has prior incarnations in MCMC convergence diagnostics (Vehtari et al, 2021) and as an additional test quantity in simulation-based calibration (Modrák et al, 2023). KS tests on PIT values are standard in econometric density-forecast evaluation (Diebold et al, 1998; Berkowitz, 2001; Czado et al, 2009; Rossi and Sekhposyan, 2019), and the Brownian-bridge covariance structure that motivates them is classical (Shorack and Wellner, 1986). Equivalence testing in biostatistics (Schuirmann, 1987; Lakens, 2017) underwrites the tolerance-band null we adopt, and sequentially valid calibration tests built from e-values and supermartingales (Vovk and Wang, 2021; Henzi and Ziegel, 2022; Arnold et al, 2023; Ramdas et al, 2023; Casgrain et al, 2024; Grünwald et al, 2024) supply the online slot. Our contribution is neither the PIT, the KS test, the folding transformation, equivalence testing, nor the e-value construction in isolation: it is their integration into a single grammar whose slots can be swapped independently and whose recipes jointly admit the directional and tolerance refinements a safety case demands.

**Classification calibration.** Calibration testing is far more mature for classification (Naeini et al, 2015; Guo et al, 2017; Vaicenavicius et al, 2019; Widmann et al, 2019; Dimitriadis et al, 2023; Silva Filho et al, 2023), with (Lee et al, 2023) giving an optimal binary test (T-Cal). Regression is structurally different: the absence of a finite label space forces reasoning about full predictive distributions rather than class probabilities, and the natural test statistics (PIT, coverage curves) carry a correlation structure that our framework exploits via the Brownian-bridge reduction of Section 4.2.

## 10 Conclusion

We have presented a four-slot framework (Model & Data, Metric, Hypothesis, Testing) that organises calibration testing for safety-critical regression as a *decision* rather than a diagnostic. Every recipe drawn from the framework is evaluated along three axes that reflect the needs of such applications: *directional safety* (one-sided formulations that reject overconfidence while accepting conservatism), *tolerance* (KS bands that encode the engineering spec "accept any model whose true coverage is at most $\varepsilon_{\text{tol}}$ short of nominal"), and *operationalisability* (every recipe ends in an accept/reject decision that can be written into a safety case).

The two experiments sit at opposite corners of the design space and share no slot: weather is an offline two-sided KS *p*-value on a univariate Gaussian forecaster, and the robot localizer is an online one-sided likelihood-ratio e-value monitor on half-plane projections of a particle cloud. Four appendix ablations then hold the data fixed and swap one slot at a time (metric, testing, data-model, conditional), showing that the same grammar yields operationally distinct, engineer-readable decisions on identical data. A unifying observation throughout the paper is the known equivalence between per-level coverage and PIT uniformity (Gneiting et al, 2007): the KS test on folded PIT arises as the natural dense-grid limit of correlation-aware per-level coverage testing, resolving the power overhead of Bonferroni in one stroke, and its one-sided variant is the safety-critical form that composes with tolerance bands into the deployable check of Section 5.

**Limitations.** Four assumptions bound the scope of the framework. (i) All tests assume i.i.d. validation data; in sequential settings this holds only approximately. (ii) The framework targets *marginal* probabilistic calibration; conditional calibration is addressed only coarsely via variance binning. (iii) Our experiments are low-dimensional, and the number of half-plane projections needed for adequate coverage of the unit sphere grows with dimension, so scaling to high-dimensional prediction spaces is still open. (iv) The e-value construction requires an explicit alternative $p_{\text{alt}}$, and a misspecified alternative may delay or miss detection.

**Future directions.** Four extensions follow naturally from the framework:

- **Conditional calibration beyond binning.** Replace quantile bins of $\hat{\sigma}$ with kernel-based or learned partitions so that conditional miscalibration localised to a specific region of input space can be detected without an exponential blow-up in the number of bins.

- **Sequential tests that account for temporal dependence.** The current e-value construction leans on conditional calibration; tests that account for explicit dependence structure (for instance block-martingale or HMM-aware variants) would sharpen the anytime-valid guarantees in settings where the dependence is strong.

- **Scaling the half-plane construction to higher-dimensional prediction spaces.** A linear-in-$p$ direction count is sufficient for bounded-moment distributions (Section 6.3), but directional miscalibration concentrated in a narrow cone grows progressively harder to detect as $p$ increases; adaptive direction selection is an obvious next step.

- **Integration with conformal prediction.** Conformal prediction already guarantees marginal coverage by construction, but it could benefit from the tolerance and one-sided machinery developed here, and the coverage-based half of the pipeline already applies to conformal outputs without modification.

## Appendix A. Slot-swap ablations

The two recipes of Section 8 each pick a single point in the design space. This appendix demonstrates the framework's modularity by holding the data fixed on one of the two experiments and swapping one slot at a time. Each subsection isolates a different slot of Fig. 1 and shows how the decision changes.

### Appendix A.1. Metric, hypothesis, and tolerance (weather)

The weather recipe of Section 8.1 fixes the metric slot to PIT uniformity with a two-sided KS test and no tolerance. Three alternatives are worth exposing on the same data: per-level coverage with Bonferroni (the pedagogical metric of Section 3), the folded one-sided KS of Section 5.1 (the safety-critical reading of the PIT metric), and a non-zero tolerance $\varepsilon_{\text{tol}}$ on the null (Section 5).

Table 4 cross-varies the metric (per-level coverage with Bonferroni, folded KS, unfolded KS), the hypothesis (one-sided vs. two-sided), and the tolerance ($\varepsilon_{\text{tol}} \in \{0, 0.02\}$) at $\alpha = 0.05$, $N = 365$, $K = 18$ levels. The folded-KS row shows the one-sided variant of Section 5.1, the natural safety-critical reading of the folded metric; the two-sided folded KS of Section 4.1 gives a similar answer on this dataset and is omitted to keep the table compact. Three operationally distinct readings of the same data fall out:

- The folded one-sided KS test *accepts* even at $\varepsilon_{\text{tol}} = 0$. A safety engineer learns that the prediction intervals cover at least as much mass as claimed; the KS band knows about the joint Brownian-bridge correlation (Section 4.2) and is uniformly tighter than Bonferroni in the central $\lambda$ range, leaving enough slack at the boundary to absorb the small per-level under-coverage.

- The unfolded KS test *rejects* in every column: this is the headline reading of Section 8.1, and it corresponds to the forecaster's small location bias rather than a safety-critical over-confidence.

- Coverage with Bonferroni rejects at $\varepsilon_{\text{tol}} = 0$ but accepts at $\varepsilon_{\text{tol}} = 0.02$. A system integrator reads this row and decides whether a 2% tolerance is acceptable for the deployment.

The coverage-with-tolerance variant is visualised in Fig. 17: the empirical coverage curve sits slightly below the diagonal, stays inside the $\varepsilon_{\text{tol}} = 0.02$ tolerance band at every level, but exits the strict Bonferroni band at several high-$\lambda$ levels. This is the textbook large-$N$ pathology of Section 5: exact-null tests reject operationally benign miscalibration as soon as the sample size is large enough to resolve it.

### Appendix A.2. Testing slot: online e-value on weather

The headline weather recipe uses an offline KS $p$-value. Swapping only the testing slot to the sequential likelihood-ratio e-value of Recipe 3 gives an online monitor on the *same* data, model, and per-level coverage metric. The running e-value at $\lambda = 0.9$ is shown in Fig. 18. The trace temporarily crosses the rejection threshold $1/\alpha = 20$ around $t \approx 100$–150, when a streak of under-covered predictions accumulates evidence of over-confidence, then retreats as more conforming predictions arrive. An anytime-valid online monitor flags miscalibration at the first crossing, even though the final batch-test result is ambiguous in the coverage-with-tolerance reading of Table 4. The testing slot is therefore operationally orthogonal to the metric and hypothesis: the same data can be audited retrospectively (KS $p$-value) or monitored online (e-value) with no re-derivation.

|  | **One-sided** | | **Two-sided** | |
| --- | --- | --- | --- | --- |
|  | $\varepsilon_{\text{tol}} = 0$ | $\varepsilon_{\text{tol}} = 0.02$ | $\varepsilon_{\text{tol}} = 0$ | $\varepsilon_{\text{tol}} = 0.02$ |
| Coverage + Bonferroni | reject | **pass** | reject | **pass** |
| One-sided folded KS | **pass** | **pass** | n/a | n/a |
| KS on unfolded PIT | reject | reject | reject | reject |

Table 4: **Metric, hypothesis, and tolerance ablation on the weather forecaster.** 3 metrics × 2 hypotheses × 2 tolerances at $\alpha = 0.05$, $N = 365$, $K = 18$ levels. Decisions decompose predictably across slots: the one-sided folded KS test dominates Bonferroni (accepts where Bonferroni rejects), tolerance bands rescue Bonferroni from large-$N$ rejection, and the unfolded KS (the headline metric of Section 8.1) detects the forecaster's small location bias, which the folded variant correctly absorbs.

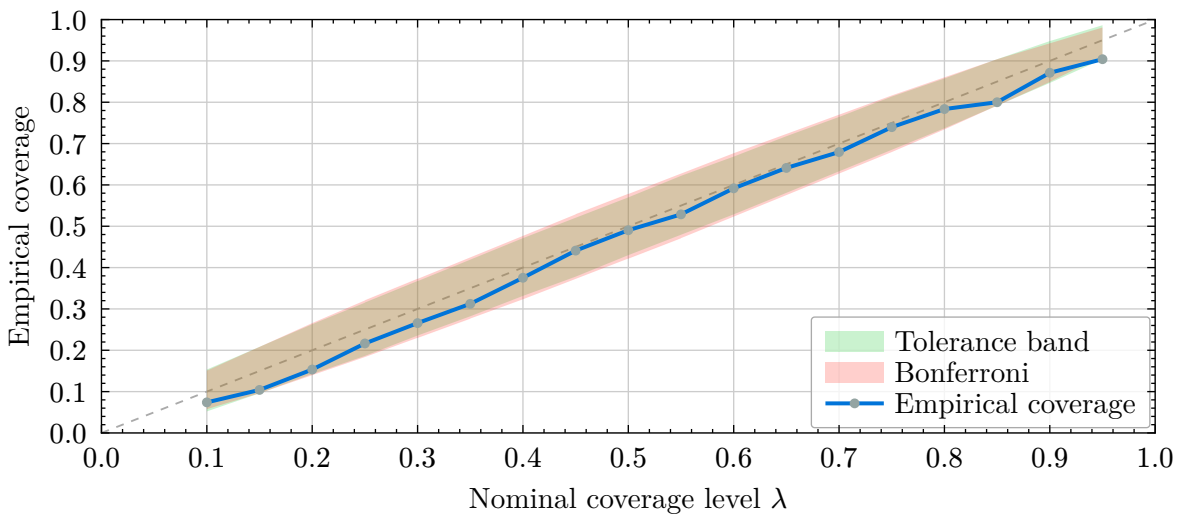

Fig. 17: **Coverage with tolerance band on the weather forecaster.** The empirical coverage stays within the tolerance band ($\varepsilon_{\text{tol}} = 0.02$, green) despite sitting below the diagonal. Without the tolerance, the strict Bonferroni bounds ($\varepsilon_{\text{tol}} = 0$) would reject at several high coverage levels.

**Appendix A.3. Data-model slot: Gaussian approximation to the particle cloud**

The robot recipe of Section 8.2 uses the particle cloud directly. The data-model slot can instead be swapped for a fitted multivariate Gaussian: at each step, moment-match the weighted particles to obtain $\mathcal{N}(\hat{\boldsymbol{\mu}}_t, \hat{\boldsymbol{\Sigma}}_t)$ and apply the multivariate Gaussian pipeline of Section 6.1, with PIT built from the squared Mahalanobis distance $D_t^2 = (\boldsymbol{y}_t - \hat{\boldsymbol{\mu}}_t)^T \hat{\boldsymbol{\Sigma}}_t^{-1} (\boldsymbol{y}_t - \hat{\boldsymbol{\mu}}_t)$ via $u_t = \text{cdf}(\chi_2^2, D_t^2)$. This discards the particle cloud's shape (for

---

1 **Input:** prediction stream $(\mu_t, \sigma_t, y_t)$ for $t = 1, 2, ...$; coverage level $\lambda$; alternative $p_{\text{alt}} < \lambda$; significance $\alpha$
2 $E_0 \leftarrow 1$
3 **for** $t = 1, 2, ...$ **do**
4 $\quad \text{lo}_t \leftarrow \text{qtl}(\mathcal{N}(\mu_t, \sigma_t^2), (1 - \lambda)/2)$
5 $\quad \text{hi}_t \leftarrow \text{qtl}(\mathcal{N}(\mu_t, \sigma_t^2), (1 + \lambda)/2)$
6 $\quad k_t \leftarrow \mathbb{1}[y_t \in [\text{lo}_t, \text{hi}_t]] \quad \triangleright$ coverage indicator
7 $\quad e_t \leftarrow (p_{\text{alt}}/\lambda)^{k_t}((1 - p_{\text{alt}})/(1 - \lambda))^{1-k_t} \quad \triangleright$ likelihood-ratio, (26)
8 $\quad E_t \leftarrow E_{t-1} \cdot e_t \quad \triangleright$ test martingale
9 $\quad$ **if** $E_t > 1/\alpha$ **then** raise alarm
10 **end**

---

Recipe 3: **Sequential e-value calibration monitor at level $\lambda$ (univariate Gaussian).** The testing slot of Recipe 1 is swapped from offline Bonferroni to a sequential test martingale (Section 7), and the hypothesis slot is realised in one-sided form via the directional alternative $p_{\text{alt}} < \lambda$. Model and metric slots are unchanged. By Ville's inequality, the alarm at the first crossing of $1/\alpha$ is anytime-valid, and $p_{\text{alt}}$ controls detection sensitivity. Used in the weather testing-slot ablation of Appendix A.2.

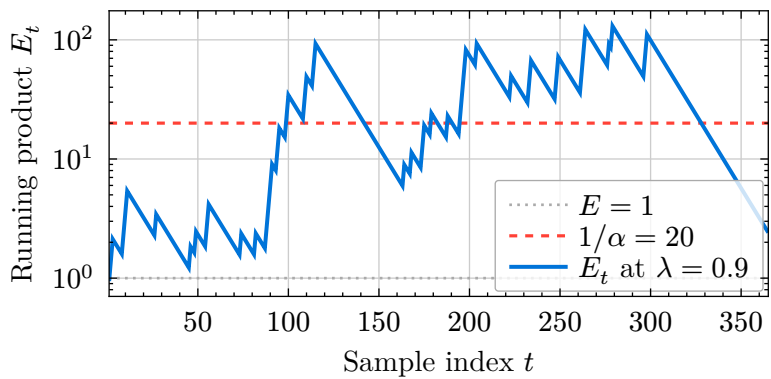

Fig. 18: **Sequential e-value trace on the weather forecaster.** Running product of per-step likelihood-ratio e-values at $\lambda = 0.9$. The threshold $1/\alpha = 20$ (dashed red) is crossed around $t \approx 100$–150, demonstrating that an online monitor would flag over-confidence even before all data is collected. This is a pure testing-slot swap relative to Section 8.1: the data, metric, hypothesis, and model are unchanged.

instance multimodality from the unmodeled drift), so mild over-confidence relative to the half-plane recipe is expected.

Fig. 20 overlays the Gaussian-approximation coverage curve on the half-plane coverage curve of Section 8.2. The Gaussian approximation under-covers at lower levels ($\hat{c}(0.5) = 0.46$ vs. 0.49 for the particles), consistent with the expected over-confidence. Fig. 21 shows the folded PIT histogram from the Mahalanobis reduction: roughly uniform, with a slight excess near $v = 1$ that signals occasional over-confidence when the drift pushes the true position into the Gaussian's tails. A Bonferroni one-sided coverage check passes for both data-model choices, but the more sensitive KS test on the Mahalanobis PIT rejects ($p = 0.046$), a subtle distributional mismatch invisible to Bonferroni and an instance of the power–interpretability trade-off of Section 4.4. Either swap is legitimate under the framework, and the choice is driven by whether the deployed monitor has access to the particle cloud (preferred) or only to its moment summary.

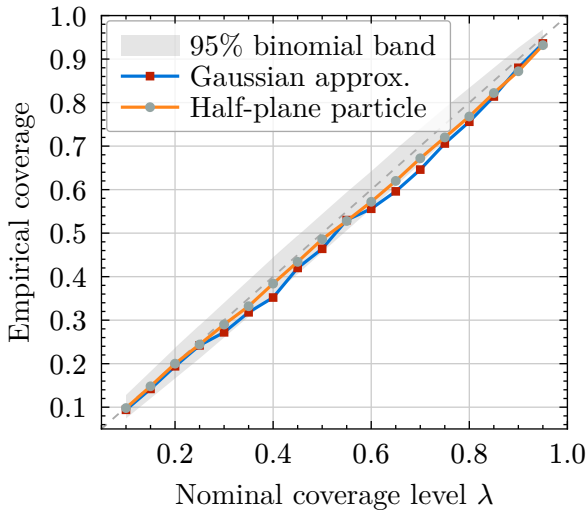

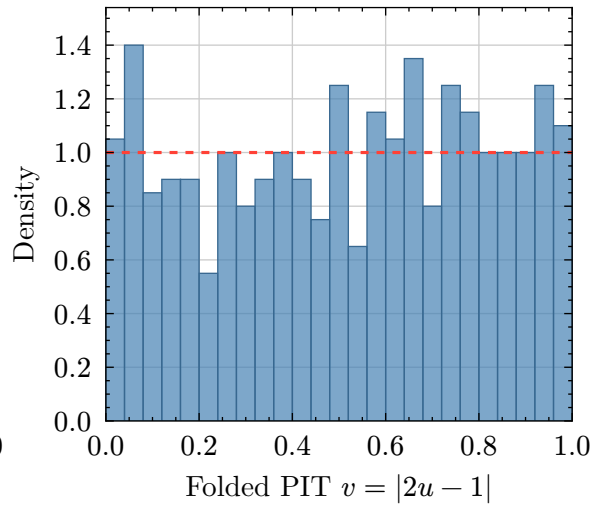

Fig. 20: **Coverage comparison.** Gaussian approximation (blue) and half-plane particles (orange) on the same trajectory. Both stay within the 95% Binomial band, but the Gaussian curve sits slightly lower, reflecting mild over-confidence from discarding the cloud's shape.

Fig. 21: **Folded PIT histogram from the Mahalanobis reduction.** The slight excess near $v = 1$ indicates occasional over-confidence from the unmodeled drift; the rest of the distribution is close to uniform.

### Appendix A.4. Conditional calibration: variance binning on the weather forecaster

Variance binning wraps any marginal check with conditioning on the predicted uncertainty, partitioning predictions into bins of $\hat{\sigma}$ and rerunning the per-bin check. We partition the weather predictions into three bins (low, medium, high; $\approx 122$ samples each) and rerun the two-sided tolerance check of Appendix A.1 inside each bin. All three bins pass (Fig. 22), indicating that the weather forecaster's calibration quality does not vary systematically with its own predicted confidence. With $\approx 120$ samples per bin, per-bin power is limited, but the absence of a clear conditional pattern is reassuring: a miscalibration that hides in a specific variance regime would show up here as a failing bin, and none do.

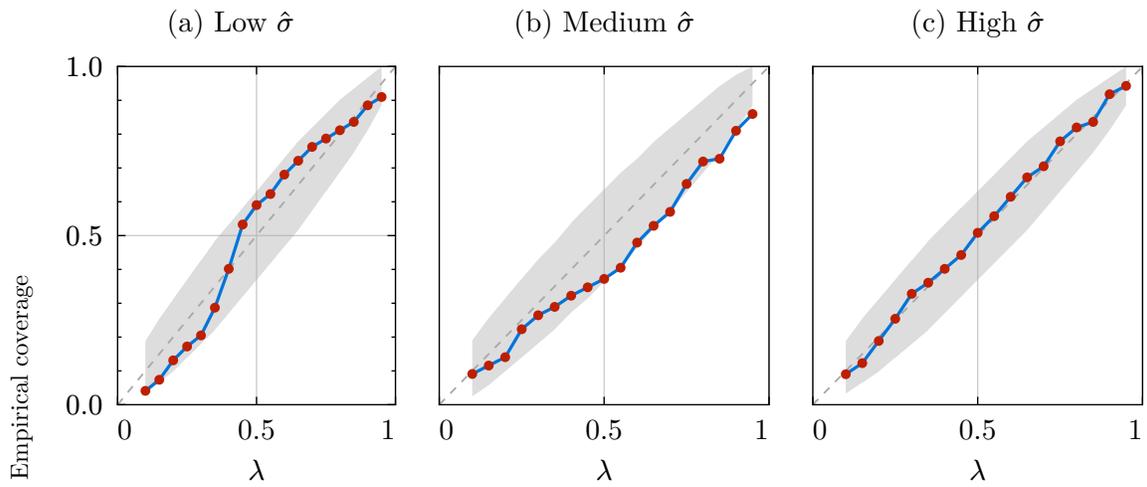

Fig. 22: **Variance binning on the weather forecaster.** Per-bin coverage curves for low, medium, and high predicted uncertainty $\hat{\sigma}$. All bins pass the calibration check, indicating no significant conditional miscalibration across the predicted-variance range.